\documentclass[%
 reprint,
superscriptaddress,
 amsmath,amssymb,
 aps,
]{revtex4-2} %

\usepackage{graphicx}
\usepackage{dcolumn}
\usepackage{bm}
\usepackage{braket}
\newcommand{\nv}{NV$^-$}
\newcommand{\nOne}{$\langle 111\rangle$}
\newcommand{\nzero}{$\langle 100\rangle$}

\newcommand{\bdc}{$\textbf{B}_{dc}$}

\usepackage{gensymb}
\begin{document}

\title{Portable diamond maser with reduced magnetic field through orientation}

\author{Wern Ng}
\email{wern.ng@imperial.ac.uk}

  \author{Yongqiang Wen}

 \author{Neil Alford}
 
 \author{Daan M. Arroo}
 \affiliation{Department of Materials, Imperial College London, South Kensington, SW7 2AZ London, United Kingdom}
\date{\today}

\begin{abstract}
    Masers have the potential to transform medical sensing and boost qubit readout detection due to their superb low-noise amplification. The negatively-charged nitrogen vacancy (\nv) diamond maser is the only continuous-wave solid-state maser discovered at room temperature, however it suffers from requiring large and bulky magnets which prevent its more widespread use. We present a significant reduction in size of the entire diamond maser using a much lighter and small-footprint electromagnet, reducing the weight from an immovable 2000 kilograms to a portable 30 kilograms. We achieve a maximum maser output power near -80 dBm, ten times higher than the first implementation, and have discovered techniques to reduce the magnetic field strength required for masing by precise manipulation of the spin orientation. With the diamond maser now shrunk to a size that can fit on a lab benchtop, we have brought continuous-wave room temperature masers away from the confines of research laboratories and closer to transforming readouts in quantum computing, frequency standards and quantum-limited medical sensing.
\end{abstract}
\maketitle
\section{Introduction}
Microwave frequencies are ubiquitous throughout medical diagnostics, space communication and mobile networks. Masers, the microwave analogues of lasers, can act as ultra-low noise amplifiers and sensors able to detect the faintest microwave signals\cite{WuHao2022,Attwood2023,Daan2021,AharonBlank2022}. Furthermore, they provide fertile ground for the study of quantum phenomena such as cavity quantum electrodynamics (cQED)\cite{Breeze2017cqed,Zhang2022,wernDAP2023} and superradiance\cite{Fokina2021,WuQilong2021,wu2022superradiant,Gottscholl2022}. However, the sole room-temperature continuous-wave solid-state maser to date, \nv\ doped diamond (henceforth denoted `\nv\ diamond'), has required large magnetic fields in excess of 400 mT to support its population inversion. All known implementations still used electromagnets that weigh almost 2000 kg\cite{Breeze2018,Zollitsch2023}. Far from being portable, the \nv\ diamond maser is still relegated to research labs, in contrast to lasers which have already been successfully minimised.

We report on a portable \nv\ diamond maser with considerable reductions to weight and footprint. The electromagnet used for supplying the magnetic field only weighs $\sim$30 kg, supports a very uniform magnetic field, and occupies a space on the benchtop no greater than a household microwave oven. We present the performance characteristics of our portable \nv\ diamond maser, and report on work towards reducing the magnetic field constraints of maser operation so as to encourage the use of even more compact magnets.
\section{Results}
\subsection{Electromagnet setup and resonator simulation}\label{subsection:setup}
\begin{figure*}
\centering
\includegraphics[width=\textwidth]{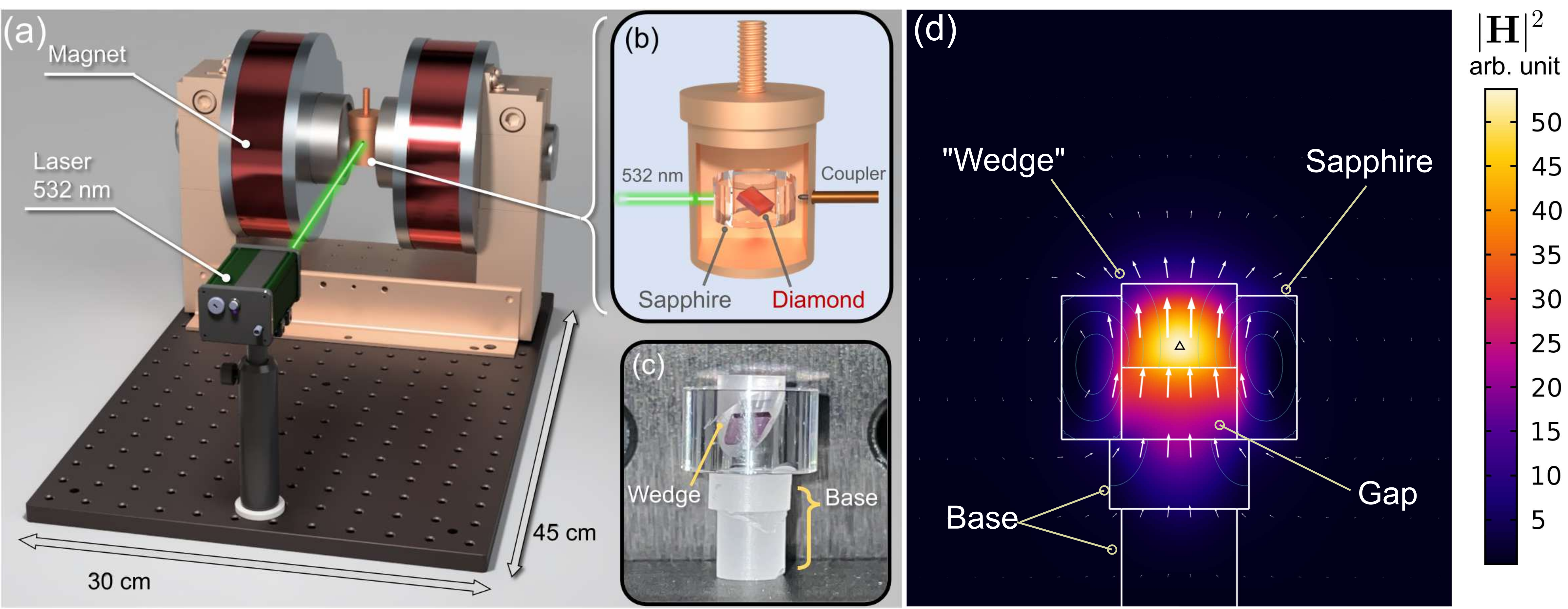}
\caption{\label{fig:magnet}(a) Diagram of the electromagnet and resonator with laser excitation. The entire setup can fit onto a standard optical breadboard (45 $\times$ 30 $\times$ 1.27 cm) such as displayed. (b) Diagram showing the inside of the maser resonator, where the diamond is inserted within the sapphire dielectric resonator. (c) Photo of the sapphire resonator and diamond outside the copper cavity, where the wedge holding the diamond can be seen. (d) COMSOL simulation of the TE$_{01\delta}$ mode of the resonator near 9648 MHz. The colour map indicates the strength of the magnetic field (as $|\textbf{H}|^2$) while the white arrows indicate the direction of magnetic field flow. The region of highest magnetic field strength is marked with a black triangle.}
\end{figure*}

The \nv\ diamond maser operates through a population inversion of the ground state triplet levels, which consist of the lowest $\ket{0}$ state and two higher energy degenerate states $\ket{\pm1}$\cite{Breeze2018}. This is achieved through firstly exciting the \nv\ diamond with 532 nm light to form a larger population in the $\ket{0}$ state through intersystem-crossing processes\cite{Breeze2018}. A dc magnetic field near 400 mT then splits the $\ket{\pm1}$ levels through the Zeeman effect, causing $\ket{-1}$ to become lower in energy than $\ket{0}$ with a lower population, which results in a population inversion between $\ket{0}$ and $\ket{-1}$\cite{Breeze2018}. Figure~\ref{fig:magnet}(a) shows the maser setup where a resonator containing the \nv\ diamond sample (the gain medium) is placed between the poles of the electromagnet. The electromagnet occupies a volume no greater than 29 $\times$ 20 $\times$ 21 cm, allowing it to be placed easily on a standard optical breadboard. Figure~\ref{fig:magnet}(b) shows a cross-section of the maser resonator consisting of an outer copper cavity housing a dielectric ring made of optically polished sapphire. 

The diamond plate sample is placed in the ring bore. A coupling loop attached to a coaxial cable inserted into the copper cavity is used to couple to the TE$_{01\delta}$ mode of the sapphire resonator and detect the maser signal. The microwave frequency of the resonator can be tuned by adjusting the inner ceiling height via a tuning screw at the top of the copper cavity. For all experiments, the resonator was tuned to 9648 MHz. Figure~\ref{fig:magnet}(c) is a photo of the sapphire dielectric outside of the copper cavity in more detail. The \nv\ diamond is affixed to a sapphire cylindrical wedge with a 45\degree\ angle, with the wedge inserted into the sapphire ring. The significance of this angle in relation to the geometry of diamond is explained in later subsections. 

The TE$_{01\delta}$ mode can be visualised through finite element simulations such as COMSOL. In Figure~\ref{fig:magnet}(d), a cross-section of the resonator is visualised through an axisymmetric approximation\cite{OxborrowAxis}. As the simulation assumes cylindrical symmetry, it cannot simulate the wedge with its axially-unsymmetric 45\degree\ slant, so the wedge is modelled flat while still filling up a portion of the hole of the sapphire ring. The region of highest magnetic field strength is shown to be at the bottom of the wedge, hence a sample placed in this region would have the highest interaction with the microwave mode which is ideal for masing. We note that this is different to previous maser work where simulations showed the magnetic field hotspots to be on the sides of the hole, when no wedge of identical material is present\cite{HaoWu2020_cont}. 

A key factor in determining the threshold optical pump power required for masing is the Purcell factor $F_P$\cite{Purcellfactor} which is directly proportional to the loaded quality factor $Q_L$ of the resonator divided by the mode volume $V_m$ of the resonator ($F_P\propto Q_L/V_M$). Hence, it is useful to characterise these two values for the resonator; $Q_L$ is measurable using a vector network analyser (VNA), while the mode volume can be calculated using COMSOL, which gave a value of 0.18 cm$^{3}$. Refer to the Supplementary Information for the formula of the mode volume.

For masing, the \nv\ diamond is usually oriented with its \nOne\ axis as parallel as possible with the dc magnetic field $\textbf{B}_{dc}$\cite{Breeze2018}. For our experiments, alignment was carried out manually by eye. The diamond plate is cut such that its \nzero\ axis is a vector perpendicular to the largest flat faces, so aligning \nOne\ along $\textbf{B}_{dc}$ proceeds as follows: firstly, the largest flat face of the \nv\ diamond is attached to the wedge, parallel to the slant of the wedge. Secondly, the wedge is rotated such that the slant face is facing perpendicular to $\textbf{B}_{dc}$. In Figure~\ref{fig:magnet}(a), this second step causes the slant face of the wedge to face the laser beam. Finally, rotate the entire wedge 45\degree\ along its cylindrical axis. The wedge face is then no longer facing the laser beam, but is instead 45\degree\ offset. Figure~\ref{fig:magnet}(c) gives an approximate idea of this offset if looking straight at the \nv\ diamond from the laser's perspective (though in this image, the rotation is not exactly 45\degree). Although this operation does not exactly align the \nOne\ axis to $\textbf{B}_{dc}$, this alignment is sufficient for masing as will be seen later with measured signals.

The limit by which the \nOne\ axis can be misaligned from $\textbf{B}_{dc}$ while still maintaining masing capability is an intriguing parameter that will be explored in a later subsection, as it would not only help benchmark the margin for error for maser operation, but also allow operation with lower magnetic field requirements.
\begin{figure*}
\centering
\includegraphics[width=1.\textwidth]{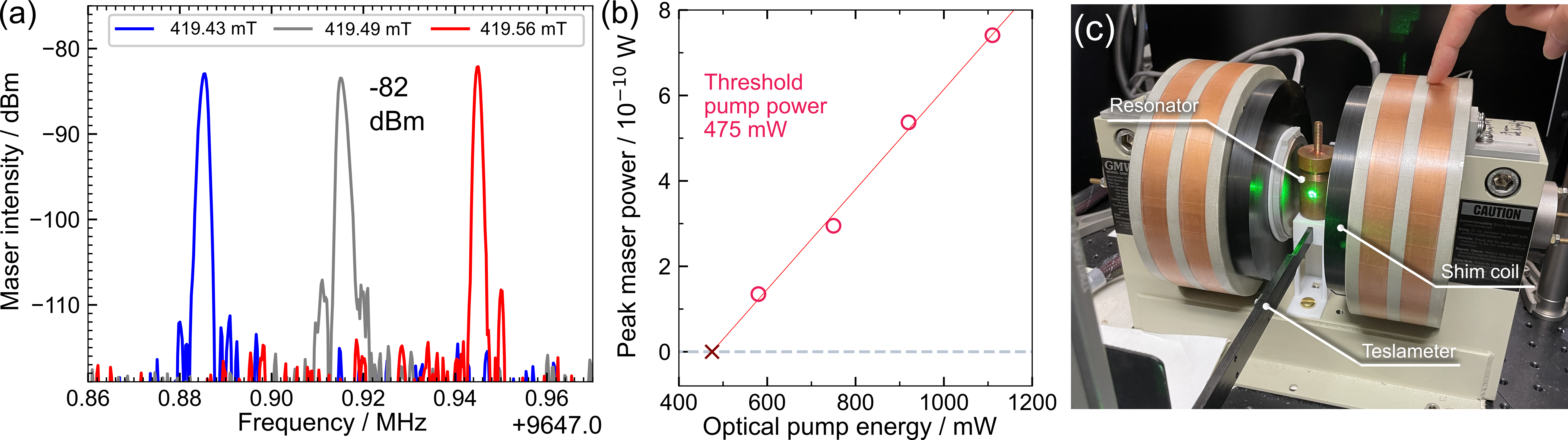}
\caption{\label{fig:mainsignal}(a) Single shot captures of masing signals at each hyperfine resonance for \nv\ diamond. All signals share a similar amplitude of approximately -82 dBm, and the magnetic fields at which the signals were detected are indicated. (b) Threshold laser power required for masing, calculated using different excitation powers. (c) Photo of the maser cavity in the electromagnet under laser excitation.}
\end{figure*}
\subsection{Masing signals}
Figure~\ref{fig:mainsignal}(a) shows the output masing signals (with three peaks corresponding to three 14N hyperfine transitions) for various applied magnetic fields, and Figure~\ref{fig:mainsignal}(c) shows a photo of the maser resonator inside the poles of the electromagnet being excited by the 532 nm laser at 1570 mW. These signals were measured at a particular orientation of the \nv\ diamond (discussed in a later subsection), and gave peak strengths of approximately -82 dBm. This is stronger than the first iteration of the \nv\ diamond maser at -90 dBm\cite{Breeze2018}, and though the masing amplitude reported here is not as strong as that of the most recent work in literature\cite{Zollitsch2023}, our maser operates at a fraction of the weight and footprint of that work. The hyperfines have a microwave frequency spacing of $\sim$30 kHz, with a magnetic field spacing of 0.06 mT (0.6 Gauss) between the left and centre hyperfine and 0.07 mT between the centre and right hyperfine. The magnetic field positions were 419.43, 419.49 and 419.56 mT from the left to the right-most hyperfine respectively, with an error of $\pm$0.03 mT. The hyperfine spacing is similar to that reported in EPR measurements in previous literature at 9570.5 MHz\cite{Breeze2018}, which report a spacing of approximately 0.075 mT between two hyperfines. Given the resonator has a $Q_L=21800$, the bandwidth would be 440 kHz which allows it to encompass all three hyperfines.

Figure~\ref{fig:mainsignal}(b) shows the peak maser power varying with laser excitation, where a maser was detectable down to less than 600 mW of input laser excitation. This corresponds to a threshold of 475 mW for the maser, under the same $Q_L$ and mode volume. Furthermore, it was observed that masing was still possible up to a power of 2200 mW. However, near 3700 mW the maser signal disappears and increasing $Q_L$ by decreasing the coupling did not help to revive the maser at this pump power. It is known that at very high laser powers, the \nv\ centres begin to convert to NV$^{0}$ centres\cite{Aslam_2013} which cannot mase, and so this may be the cause for the loss of a maser signal. This process is not permanent though; using a lower laser power again, such as 1570 mW, will give a maser signal.

\subsection{\nv\ orientation control}\label{sub:orientationmethod}

Would there be an impact on masing if \bdc\ was not aligned parallel to the \nv\ centre? Na\"{i}vely, one may not think so since this should just change the energy level splitting between $\ket{0}$ and $\ket{-1}$. As long as an orientation gives an energy splitting with a population inversion, then masing should be possible. This would be attractive for reducing magnetic field requirements while preserving the same masing frequency, and help with probing the extent to which an error in alignment could prevent masing. A small angle misalignment of \nOne\ from \bdc\ results in the energy splitting between $\ket{0}$ and $\ket{-1}$ increasing, meaning for the same microwave frequency, the resonance would occur at a \textit{lower} magnetic field strength, thus reducing the magnetic field needed to mase. However, previous studies have shown that $T_1$ and $T_2$ are both affected by orientation, where not aligning parallel to \bdc\ results in a decrease for both lifetimes\cite{KOLLARICS2022}. The lifetimes are then anisotropic, and in order to mase above threshold, these two lifetimes should ideally be as high as possible (giving a higher cooperativity). Hence, the masing performance should degrade the more misalignment there is, but to what extent?

Before proceeding, we evaluate the magnetic field for to the maximum splitting when perfectly aligning \nOne\ to \bdc. The resonant frequency was approximately 9648 MHz for all measurements. Using the equation for the Zeeman splitting of triplet electrons where $m_s=\pm1/2$, the central magnetic field resonance can be found using the equation $B=\Delta E/(g\mu_B)=hf/(g\mu_B)$ where $h$ is Planck's constant, $f$ is the resonance at 9648 MHz, $g=2.0023$ for the free electron and $\mu_B$ is the Bohr magneton. This gives a value of 344.27 mT at the centre resonance. In previous literature that measured electron paramagnetic resonance (EPR) of the triplets in \nv\ diamond at a microwave frequency of 9570.5 MHz, the triplet resonances for the maximum splitting occurred at 239 mT and 444 mT\cite{Breeze2018}. This splitting of 205 mT scales with the microwave frequency, and so at 9648 MHz the resonances of maximum splitting would be similar at 241 mT and 448 mT.
\begin{figure*}
\centering
\includegraphics[width=0.95\textwidth]{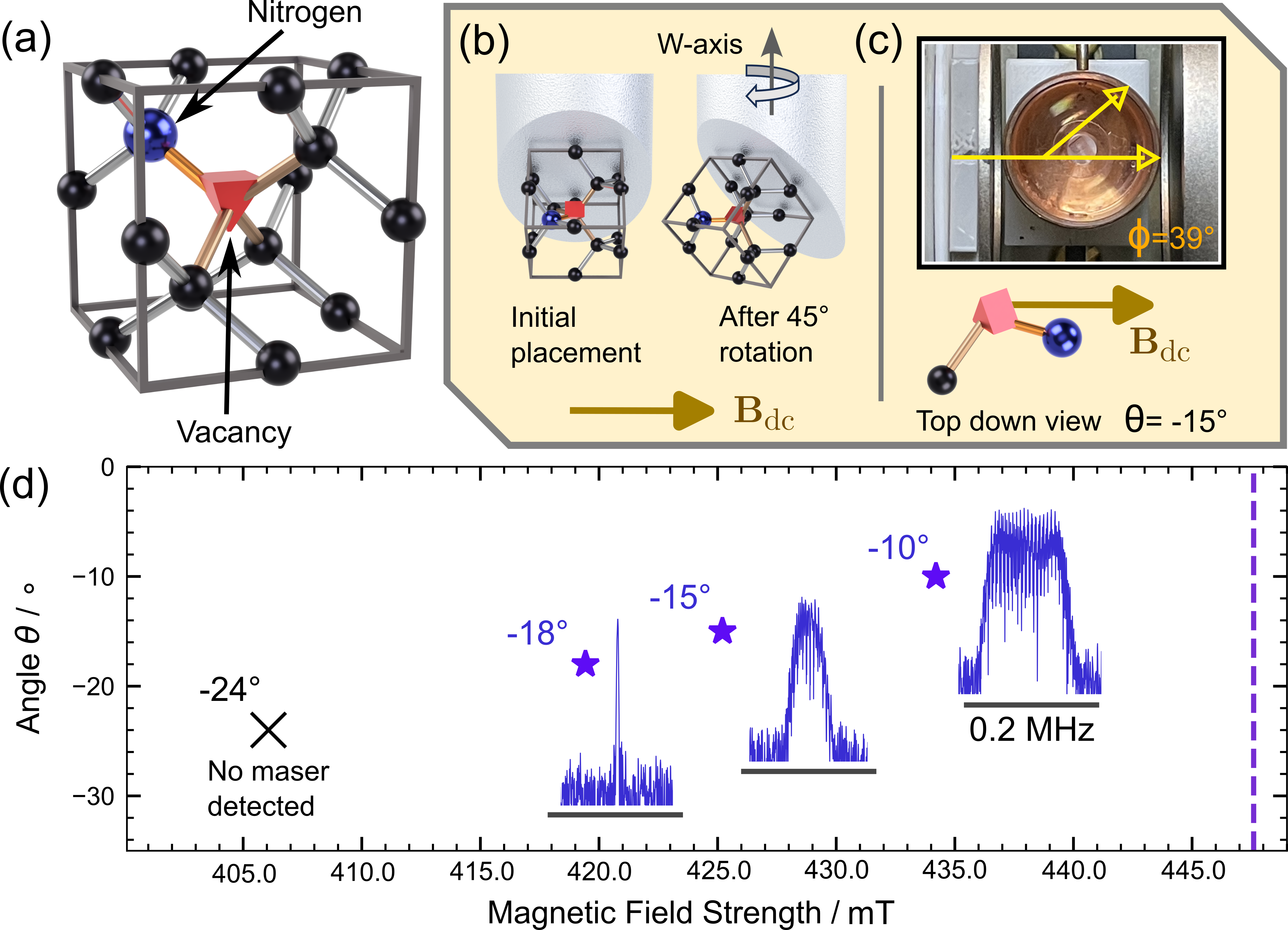}
\caption{\label{fig:ori}(a) Diamond cubic lattice cell with carbon atoms in black and the nitrogen with vacancy labelled. The golden `bond' between the nitrogen and vacancy defines the \nv\ vector. (b) Render of the rotation performed on the diamond when placed on the wedge as in Figure~\ref{fig:magnet}(c). The view is from the laser's point of view, and the \nv\ vector can be seen to be nearly parallel with \bdc. (c) View of the diamond and wedge in the resonator with copper cap removed. The top image yellow arrows define $\phi$ through the visible edge of the diamond plate, while the bottom image defines $\theta$. (d) Magnetic field position of the masing signal with varying $\theta$. The measured single shot masing signal in frequency is shown next to each star marker, with a scale bar indicating a width of 0.2 MHz. A dashed line indicates the magnetic field if the \nv\ vector was parallel ($\theta=0$)}
\end{figure*}

Next, we briefly map out the geometry of the \nv\ centre. The crystal lattice cell of diamond is cubic, as shown in Figure~\ref{fig:ori}(a). In diamond, a carbon atom bonds to four others in a tetrahedral structure, and so the angle between all carbon-carbon bonds is 109.5\degree\cite{Loubser_Wyk_1978}. Forming an \nv\ center involves replacing two carbons with a nitrogen and a vacancy, and the `\nv\ vector' (spin direction) is formed by tracing from the vacancy to the nitrogen atom, which is parallel to \nOne. This tetrahedral structure also means there are four potential positions where a nitrogen atom can substitute, each forming their unique \nv\ vector that would be 109.5\degree\ from each of the other three. This explains the presence of four pairs of triplet signals in EPR of \nv\ diamond in literature due to resonances from four different \nv\ vector orientations\cite{Breeze2018}. The plate sample was cut such that the \nOne\ axis lies on the corners of the plate.

The splitting of the energy levels $\ket{+1}$ and $\ket{-1}$ relies on how parallel an \nv\ vector is to $\textbf{B}_{dc}$, i.e. if the \nv\ vector is set parallel with $\textbf{B}_{dc}$, then rotating the sample by 180\degree\ around the wedge cylindrical axis (now denoted as `W-axis') will produce the same energy splitting. This then leads to the method explained in Subsection~\ref{subsection:setup}, where the plate is first attached to the 45\degree\ slant of the wedge. This places two potential \nv\ vectors on the plane perpendicular to the W-axis. The wedge is then rotated 45\degree\ around the W-axis (Figure~\ref{fig:ori}(b)). In reality, the tetrahedral angles mean that to align along an \nv\ vector, the rotation angle should instead be approximately 54\degree\ (109.5\degree/2). However, a 45\degree\ rotation was chosen as this was easier to gauge by eye. 
\subsection{Orientation effects on masers and reduction of magnetic fields}\label{sub:orientation}

For simplicity, we concentrate here on attaining masing signals from the highest of the high field resonances. We define the angle of rotation $\phi$ to be between \bdc\ and the visible edge of the plate from a topdown view (see Figure~\ref{fig:ori}(c) for where this edge is), and four different angles were tested: $\phi=$ 44\degree, 39\degree, 36\degree\ and 30\degree. These angles were only measured after the masing experiments through measuring angles in images of the setup (for example, as in Figure~\ref{fig:ori}(c)). The angle between \nOne\ (i.e. one of the \nv\ vectors in the plane) and \bdc, which we denote $\theta$, can be found by trigonometry from subtracting 54\degree\ from $\phi$. This gives  $\theta=$ -10\degree, -15\degree, -18\degree\ and -24\degree\ between \bdc\ and one of the \nv\ vectors in the plane. If $\theta=0$, the \nv\ vector is parallel with \bdc, and the larger the absolute value of $\theta$, the larger the misalignment from being parallel. Only sample alignment was changed for these maser experiments; the optical excitation power, optical alignment, coupling loop position, $Q_L$ and resonator microwave frequency were kept the same. 

If the \nv\ vector is completely parallel with \bdc\, the resonance would be near 448 mT. However, since the two \nv\ vectors in the plane (the out-of-plane vectors are ignored for simplicity) form an angle of 109.5\degree\ with each other, then by symmetry, the wedge just needs to be rotated such that $\theta=-35.25$\degree\ (i.e. [109.5\degree - 180\degree]/2) before it begins to increase in overlap with the other \nv\ vector. Figure~\ref{fig:ori}(d) shows how the masing signal (collected from the leftmost hyperfines) will shift to lower magnetic fields as $\theta$ becomes more negative (i.e. less alignment of the \nv\ vector with \bdc). All the signals maintained the same centre microwave frequency. The magnetic fields corresponding to $\theta=$ -10\degree, -15\degree, and -18\degree\ were 434.20, 425.20, and 419.43 mT respectively. This shows successful reduction of the magnetic field through spatial orientation. 

However, there is a limit to this rotation; at $\theta=-24\degree$, the masing signal is lost. The resulting magnetic field position recorded at 406.07 mT was only the position of the EPR signal (see Methods). This may be caused by the aforementioned reduction in $T_1$ and $T_2$ resulting from the misalignment, which would eventually bring the maser below threshold and extinguish it. Hence, the most one can stray from alignment would be $\theta=-18\degree$ while still maintaining a relatively healthy maser signal, given the current laser power and sample properties. 

The three orientations which successfully gave a maser signal have the corresponding signal trace also displayed in Figure~\ref{fig:ori}(d). All maser signals had similar amplitudes between -85 and -82 dBm. The angle of $\theta=-18\degree$ was the orientation by which the signals in Figure~\ref{fig:mainsignal}(a) were collected, with the left hyperfine signal being displayed in Figure~\ref{fig:ori}(d). Underneath each signal is a scale bar 0.2 MHz in length, and it can be seen that as the \nv\ vector is more aligned ($\theta$ approaches zero), the maser signals are subject to a similar limit-cycle broadening as seen in previous work\cite{Breeze2018}, with the lines broadening to just below 0.2 MHz at $\theta=-10\degree$. The larger this broadening, the higher the diamond is above the masing threshold, and so this indicates that the diamond would be easier to mase as the \nv\ vector becomes more aligned to \bdc. 

If one is high above the masing threshold, one can increase the amplitude of the maser and narrow its bandwidth by increasing the microwave coupling. This would reduce the $Q_L$ which would make it harder to mase, but this could be compensated by aligning closer to \bdc. It has been shown that masing with diamond requires a complex interplay of parameters that include the resonator, light pumping and even sample concentration dependence\cite{Yongqiang2023,Zollitsch2023}, but it is possible to compensate for a weaker parameter by increasing the strength of others. Hence, if the misalignment of the \nv\ vector from \bdc\ only causes $T_1$ and $T_2$ to decrease, then this may be compensated by increasing other parameters of the maser to increase the cooperativity above threshold. For example, larger pump powers or a more concentrated \nv\ diamond (while avoiding $T_2$ reducing too much) could allow masing to be possible at angles farther from \bdc, and hence achieve lower magnetic field requirements. 

However, with the current sample there is already a generous angular range the diamond can be placed while still being able to mase, which would be $\pm$18\degree\ around \bdc\ (a total range of 36\degree). The reduction of magnetic field from the maximum splitting is also significant at almost 30 mT from the maximum splitting (447 mT), which gives the diamond maser much more flexibility in its operation.

\section{Discussion}

We have demonstrated an \nv\ diamond maser at a significantly reduced form factor, forgoing the 2000 kg electromagnet of previous implementations with a much lighter 30 kg system, while still surpassing the output power of that in the first implementation of the diamond maser\cite{Breeze2018}. Through manipulation of the diamond orientation, we were able to operate the maser at a range of magnetic field strengths (magnetic tuning) without needing stringent alignment to \bdc. The method we present here for rotating the diamond's orientation would be useful for reducing magnetic field requirements which could allow more compact magnet systems to be used for masing, leading to even lighter and smaller masers. This is because it is easier to make more uniform and smaller magnet systems if the field strengths required are not too high.

The portable maser could be improved in a few aspects; the maser signals shift slightly in frequency with time (see Supplementary Information). We believe the source of this shifting is noise from the electromagnet power supply, which could be solved using a more current-stable power supply. Our future work aims to forgo such a power supply altogether and utilise a Halbach array of permanent magnets, which could completely eliminate the fast varying current noise which magnet power supplies suffer from. This would also further shrink the footprint of the diamond maser, where it may be comparable to the size of the portable pulsed pentacene maser in our previous work\cite{Wern2024}. This benchtop diamond maser has been an important stepping stone towards achieving this goal, as we have shown that large magnets are not necessary to produce \nv\ diamond masers.

Finally, in keeping with the spirit of encouraging widespread adoption of these diamond masers, the following are some guidelines to make it easier to achieve maser signals with \nv\ diamond, as it is important to know in what aspects the maser is forgiving, and what aspects require meticulous control. Similar to previous work\cite{Zollitsch2023}, many of these parameters are dependent on sample concentration and lifetimes, and some paramaters here may not work for samples with lower \nv\ concentrations (or even over-concentrated samples). The sample used here has an \nv\ concentration of 4.5 ppm and carbon isotopic purity to help maximise $T_2$. Masing above threshold requires maximising the Purcell factor which is proportional to $Q_L/V_{mode}$. Though a $Q_L$ of 21800 was used for all experiments, this was after it was known the system could mase, and it is advised to go higher with at least $Q_L=25000$ (the higher the better). This must be balanced by the coupling though, as undercoupling to increase $Q_L$ may also extinguish the maser, and so practically a $Q_L=35000$ can be considered to be more than adequate. Stronger laser powers would also be very beneficial, where 1500 to 2500 mW would improve the chance of detecting a maser signal (i.e. use laser excitation powers well above the threshold). However, the authors do not suggest going much higher than this as the aforementioned conversion to NV$^0$ centres prevents masing. Lastly on the spatial alignment, the diamond maser is found to be quite forgiving; a misalignment of $\pm$18\degree\ around \bdc\ will still allow masing.

\section{Methods}
\label{Sec:Methods}

\subsection{Diamond sample}
The diamond sample (labelled 2070512-09, nicknamed `09') was supplied by Element Six and had dimensions of 0.50 $\times$ 2.83 $\times$ 2.89 mm. It had a high isotopic purity with 99.999\% $^{12}$C. It contained 16 ppm starting [N] with subsequent irradiation and annealing to approximately 4.5 ppm [NV]. The sample has been polished on all sides to remove any graphitic carbon. It was found that without having polished away this graphitic carbon, the sample causes the $Q_L$ of the cavity to drop to below 9000 when placed inside the sapphire resonator, where this $Q_L$ is too low for masing.

\subsection{Resonator and coupling}
The cylindrical cavity enclosing the sapphire is made from CW009A copper (electronic grade). The copper cavity has an ID of 23 mm and OD of 26 mm. The inner ceiling height was adjustable through a tuning screw, and could be used to tune the resonant frequency of the entire resonator when the sapphire was placed within. The sapphire dielectric ring, stand and wedge are all made from sapphire. The diamond plate was attached to the wedge using polycement glue (Humbrol). The authors advise using only small amounts of glue, or else it can cause $Q_L$ to reduce. Light enters through a hole on the side of the copper cavity. The sapphire ring was made by SurfaceNet, and the wedge was made by Agate Products Limited.

The coupling loop enters through a hole drilled on the opposite side of the light entry hole. The loop itself is made from the open end of a rigid coaxial cable; the inner wire of the coaxial is curved and soldered onto the outer copper coating of the coaxial, forming a loop of roughly 2 mm diameter. It is important that the coaxial cable and loop are not magnetic (at least for sections which are entering between the magnet poles). The insertion of the loop into the cavity was controlled by a manual gear mechanism, and the deeper the loop was inserted, the stronger the microwave coupling and the lower the $Q_L$. Images of the coupling loop are in the Supplementary Information

After being adjusted to an ideal position, the coupling loop was no longer moved during all experiments, and the inner cavity ceiling height was always adjusted to maintain a resonant frequency at 9648 MHz, and so the $Q_L$ remained virtually constant for all experiments. The copper cavity is screwed onto a 3D printed PLA plastic stand (visible in Figure~\ref{fig:mainsignal}(c)) that is screwed onto the electromagnet, with the stand height designed to have the diamond sample at the central height between the poles. All screws were non-magnetic brass.

\subsection{Electromagnet and power supply}

The electromagnet used was a GMW Magnet Systems Model 3840 Electromagnet with specially designed poles to maximise the uniformity of the magnetic field. The pole diameter was 45 mm and the pole gap was 30 mm.The power supply for the electromagnet was a Sorensen DLM80-37E (current to reach 440 mT was approximately 18 A), and a pair of bias coils helped provide finer tuning of the dc magnetic field. The bias coils were controlled by adjusting dc current using a Kepco Bipolar power supply (BOP 20-5DL), with the current varying between -0.5 and 0.5 A. It is noted that operating in constant current mode (for both power supplies) is essential for maintaining a more stable magnetic field, since constant voltage mode results in more unpredictable drift.

For all experiments in this paper, a Thermo Scientific ThermoFlex2500 chiller was used to provide water cooling to the electromagnet. However, it was found that smaller units such as an Applied Thermal Control (ATC) Mini Chiller (500W) were also capable of water cooling without large overheating of the magnet and still providing maser signals. Maser signals could also be observed without water cooling the electromagnet at all, but due to overheating from high currents this could not be done for more than ten minutes.

\subsection{Magnetic field characterisation}
The magnetic field strength was measured using a DTM 151 digital teslameter with digital filtering turned off. The probe (which can be seen in Figure~\ref{fig:mainsignal}(c)) was placed such that it was in the centre between the poles, but offset in height below the resonator. The offset from the centre height between the poles was 28.5 mm downwards. Through taking multiple measurements at the centre height of the poles and comparing with measurements when at the offset height, an equation relating the magnetic field at the offset height to the central height could be obtained. This equation was found to be
\begin{equation}
    B_{samp}=(B_{low}-24.468)\times1.01886+26.83
\end{equation}
$B_{samp}$ and $B_{low}$ correspond to magnetic fields at the central height and offset height respectively. The equation uses the units of Gauss. The errors in the magnetic field measurements were taken as the standard deviation of fifteen random measurements of the field at that hyperfine with the highest standard deviation chosen, and subsequently rounded up to the nearest 0.01 mT.
\subsection{Optical pumping}
532 nm CW laser excitation was provided by a `finesse 532 pure' laser from Laser Quantum. Apart from the threshold laser power experiments, the laser power into the cavity was kept constant at 1570 mW and had a beam width of 2.5 mm

\subsection{Measuring resonance through VNA}
The VNA used was an M5180 2-port analyser from Copper Mountain, which was used without any further calibration apart from its default settings. The VNA was connected to the coupling loop in the cavity and the resulting S$_{11}$ dip measured could be used to attain the resonant frequency. The $Q_L$ from an S$_{11}$ measurement (measured to be 21800) can be calculated using the `Q-circle' method which uses polar charts from a VNA\cite{kajfez1995}, and example workings have been given in our previous work\cite{wernDAP2023}. Through the polar chart, the coupling was found to be undercoupled. 

The VNA can be used to perform a rapid but less sensitive form of EPR, which was very helpful in finding the microwave resonance of the diamond using the same VNA connected directly to the resonator, without any extra equipment. If the magnetic field is scanned through a resonance, one can observe the S$_{11}$ dip on the VNA `wobble' when passing through the resonance. This was used to find the approximate resonance location for the $\theta=-24\degree$ orientation in Subsection~\ref{sub:orientation}. If a teslameter is used to measure the magnetic field at which this wobble is observed, the magnetic field positions of these resonances can be quickly obtained. This was very helpful in attaining an approximate 1 mT range within which a maser signal could be searched for in magnetic field.

This wobble is simply the direct observation of the change in $Q$ which is fundamental to observing signals in EPR\cite{RINARD1999,QuantitativeEPRHandout}, but in this case due to the long lifetime and strength of diamond EPR signals this change in $Q$ can be observed directly on the VNA without additional amplification.

\subsection{Detection equipment}
The maser signal is detected using a R5550 Real-time spectrum analyser (thinkRF). To facilitate seamless switching between monitoring the cavity resonance on the VNA and then measuring a maser signal, the coupling loop coaxial cable is connected to microwave switch (HP 8761A, battery powered) that will switch between connecting the coupling loop to the VNA or the spectrum analyser. When using the spectrum analyser, the VNA rf power output is turned off to avoid spurious signals from the VNA being detected by the spectrum analyser.

When saving data signals, the span was set to 0.5 MHz, with resolution bandwidth (RBW) set to 1 kHz. However, if searching for signals for the first time, a wider span of 10 MHz is recommended.

For Figure~\ref{fig:mainsignal}(b), the peak values were taken as the average of six maser signal peaks from the left hyperfine signal at the indicated excitation power. 

\subsection{Orientation experiments}
The method used to orient the diamond was explained in Subsection~\ref{sub:orientationmethod}. The base, sapphire ring and diamond were all attached to each other as a single body with the same polycement glue used (the wedge was inserted into the sapphire without glue, friction fit), and so changing the orientation was simply a matter of rotating this entire body around the W-axis. As this body was almost entirely cylindrically symmetric, the $Q_L$ would remain the same, and the ceiling height of the copper cavity was adjusted to the frequency of interest.

The angle of orientation for $\phi$ was obtained through photographing the top of the resonator (copper lid off, with camera level) and then measuring angles through software (ImageJ).

\section*{Acknowledgements}
This work was supported by the U.K. Engineering and Physical Sciences Research Council through Grant No. EP/V001914/1. We thank Hao Wu (Eric) for helpful advice on the manuscript and on the graphitic carbon reducing $Q_L$.

\section*{Supplementary Information} 

Additional guidelines on finding masing signals, images of the coupling loop used, frequency shifting, mode volume and magnetic field hotspot comparisons, and cavity drawings.

\newpage
\bibliography{Atemplate}

\end{document}


\title{Supplementary Information\\Portable diamond maser with reduced magnetic field through orientation}

\author{Wern Ng}
\email{wern.ng@imperial.ac.uk}

  \author{Yongqiang Wen}

 \author{Neil Alford}
 
 \author{Daan Arroo}
 \affiliation{Department of Materials, Imperial College London, South Kensington, SW7 2AZ London, United Kingdom}
\date{\today}

\maketitle

\section{Maser signal detection guidelines}

Due to slight difficulties in finding a diamond maser signal for the first time, the following are some guidelines to help future experimenters search for a signal.
\begin{enumerate}
    \item Check the inside of the copper cavity is clean from dirt/oxidation as much as possible. The presence of black-ish marks is likely a sign of oxidation which will reduce $Q_L$ significantly. To clean a copper cavity from this dirt, metal polish (such as Brasso) followed by extensive rinsing with isopropyl alcohol and then drying would help. Check $Q_L$ afterwards.
    \item Insert the coaxial coupling loop into the resonator while ensuring the loop is oriented such that it is `flat', with its plane parallel with the cavity floor (refer to Figure S~\ref{fig:couplingloop}(b)). Furthermore, the insertion of the loop into the cavity should be straight through; avoid inserting the cable at crooked angles or else the S$_{11}$ coupling becomes distorted. 
    \item Ensure that the diamond sample has a good alignment of the \nOne\ axis to the magnetic field, and that it is well illuminated by the laser. Through appropriate laser safety goggles (e.g. LG10 goggles from Thorlabs), the \nv\ diamond will be seen to glow, which can be used to check if the laser is not missing it. During excitation, the cavity resonance will shift due to heating of the resonator (for sapphire resonators, the frequency shifts downwards with higher temperature).

\end{enumerate}

\section{Coupling loop}

\begin{figure*}[ht!]
\centering
\includegraphics[width=\textwidth]{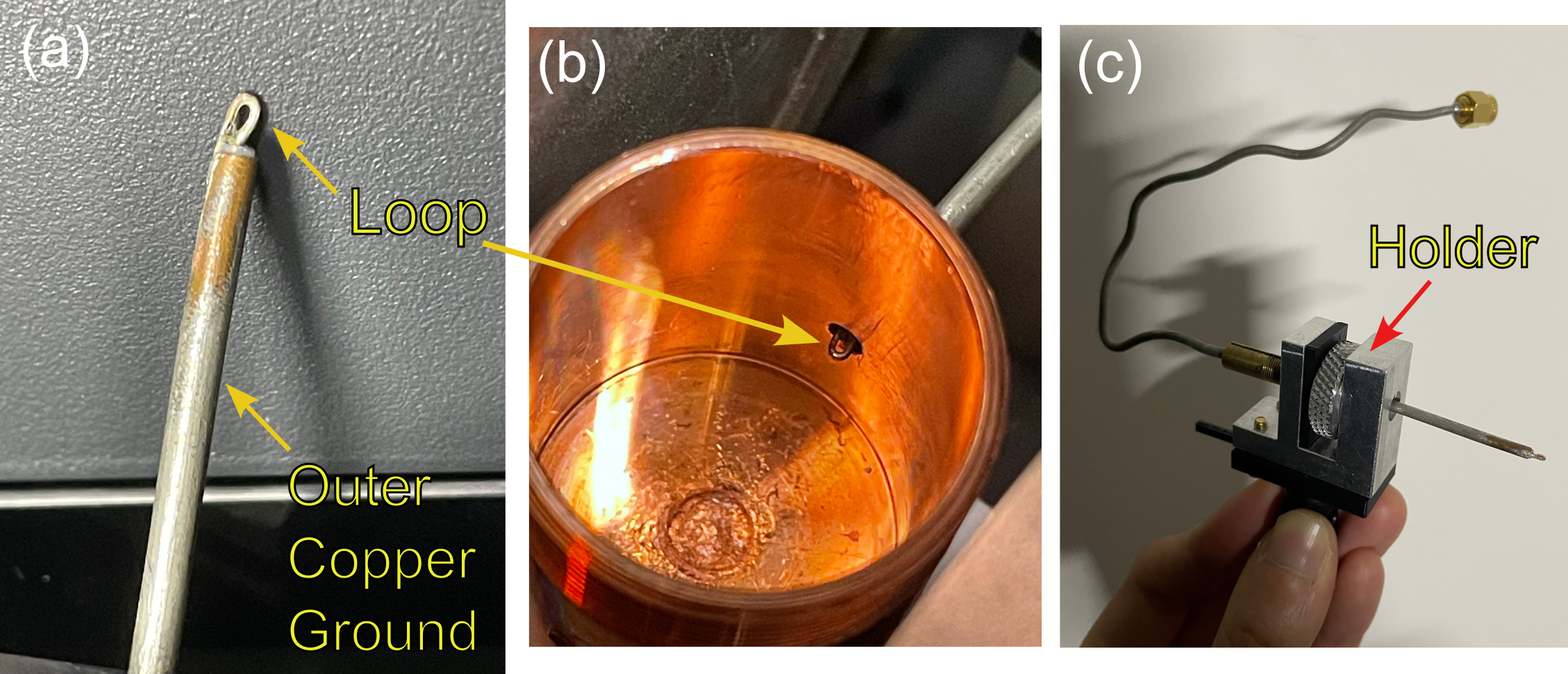}
\caption{\label{fig:couplingloop}(a) Closeup photo of coupling loop. (b) Coupling loop inserted into the hole of the resonator (sapphire dielectric not present). (c) Image of the holder for the coupling loop coaxial cable, with SMA connector visible.}
\end{figure*}
Figure S~\ref{fig:couplingloop}(a) shows the coupling loop used for coupling the maser output from the resonator. It is made from a chopped non-magnetic coaxial cable where the inner copper wire (signal line) is bent in a loop and soldered to the outer copper covering (the ground). To have precise control over the distance by which the coupling loop was inserted into the resonator, a holder based on a gear mechanism was fabricated with a combination of aluminium, brass and plastic parts (all non-magnetic), as seen in Figure~S~\ref{fig:couplingloop}(c). A screw is attached onto the coaxial cable, and is turned by the gear which has a screw hole through its centre. Inserting the coupling loop closer to the sapphire gives higher coupling and higher masing amplitudes when above threshold, but lower $Q_L$. Conversely, retracting will give lower coupling and lower masing amplitudes, but higher $Q_L$ to allow masing above threshold. Hence, if comfortably above threshold, one would try and maximise coupling to get as much masing amplitude as possible without going below threshold.
\begin{figure*}[h]
\centering
\includegraphics[width=0.9\textwidth]{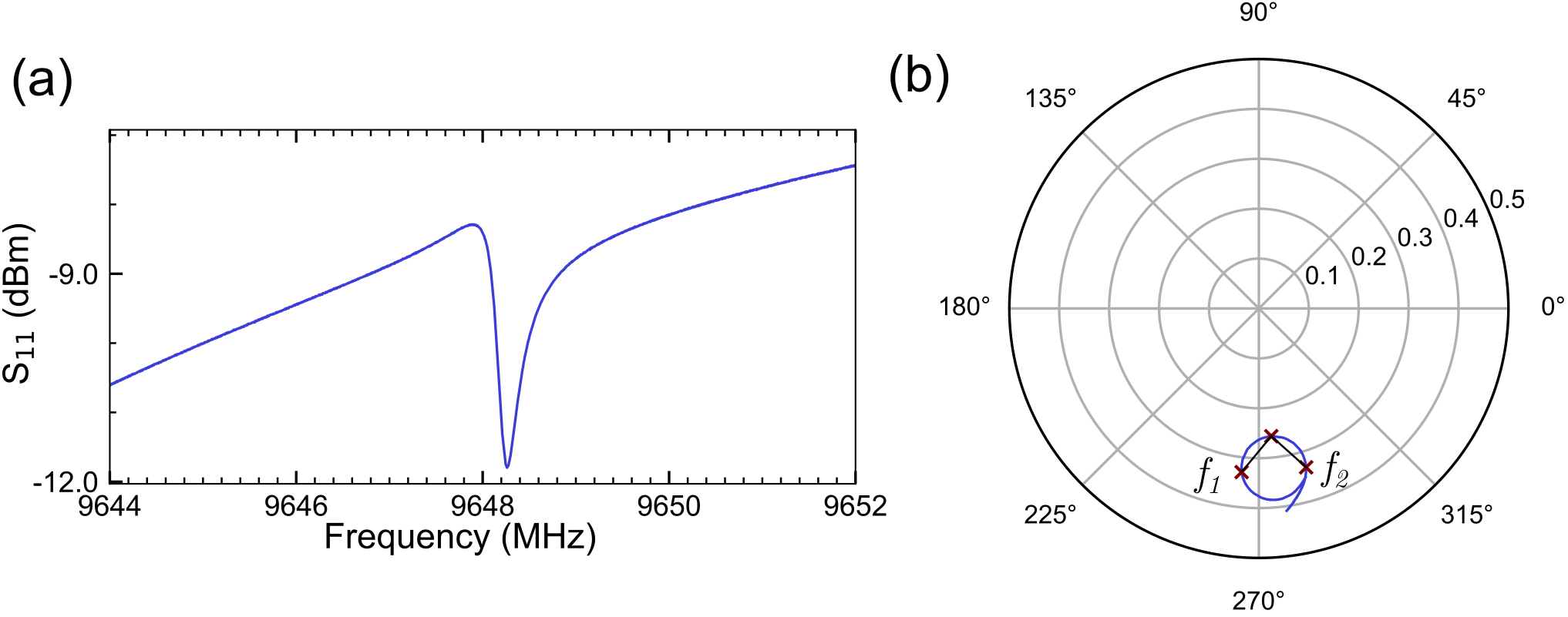}
\caption{\label{fig:VNA}(a) S$_{11}$ dip of the cavity resonance when connecting the coupling loop to the VNA. (b) Polar chart showing the $Q$-circle of the same resonance in (a)}
\end{figure*}

Figure~S~\ref{fig:VNA}(a) shows the S$_{11}$ dip from the cavity measured through the coupling loop, and the corresponding polar plot (the polar plot was \textbf{magnified} such that the radius is 0.5, since the $Q_L$-circle was very small). The $Q_L$ was calculated using the $Q$-circle method from the polar plot\cite{kajfez1995,wernDAP2023}, where there are crosses indicating $f_1$ and $f_2$ which denote the cavity bandwidth. This was calculated to be $f_2-f_1=440$ kHz. We note that though the S$_{11}$ dip appears asymmetric, this did not have a noticeable effect on masing (as long as it was not excessively asymmetric or distorted).

\section{Frequency shifting}
\begin{figure*}[h]
\centering
\includegraphics{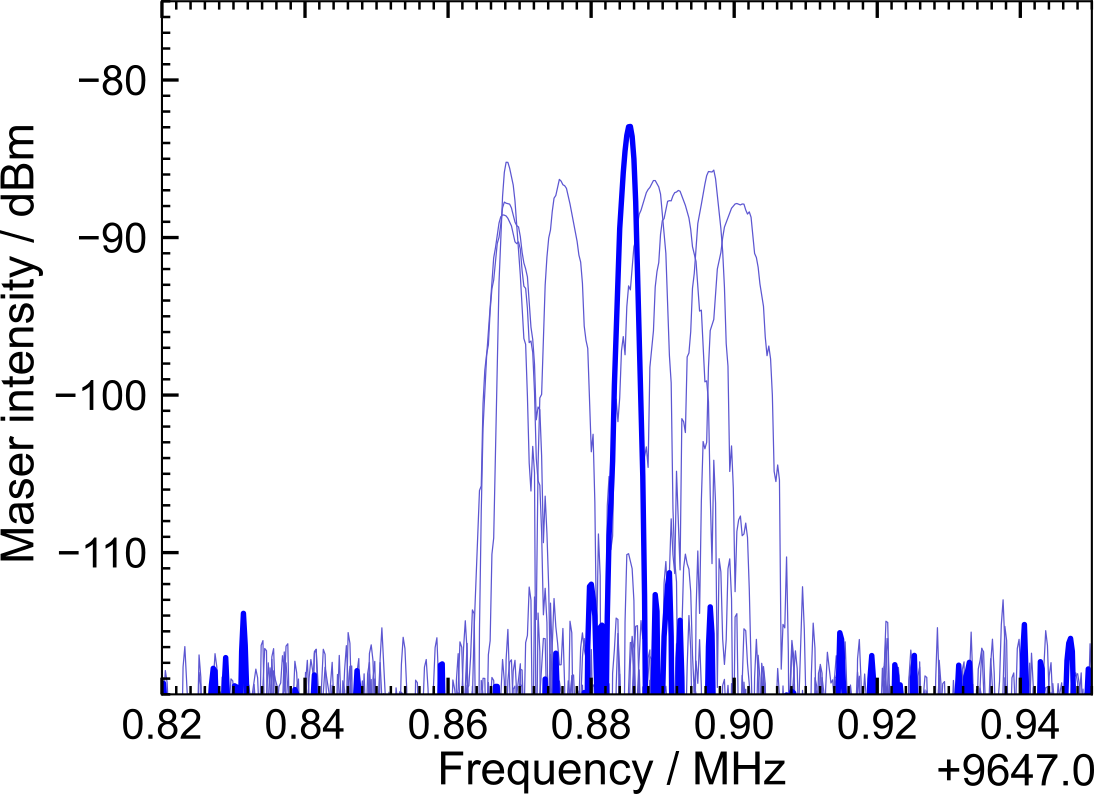}
\caption{\label{fig:jitter}Plot of nine separate single shot recordings of the masing signal}
\end{figure*}
Figure~S~\ref{fig:jitter} shows nine separate single shot masing signals collected at 419.43 mT (Figure~2(a) in the main text), corresponding to the left hyperfine. The signal corresponding to the resonance of the hyperfine was determined from the signal closest to centre of the multiple shots, which is highlighted in a thicker line and was used in Figure~2(a) of the main text. This method was similarly used to determine the signals and amplitudes of the other hyperfines.

\section{Comparison of magnetic field distribution and filling factor for resonators with and without wedge}

\begin{figure*}[h]
\centering
\includegraphics[width=\textwidth]{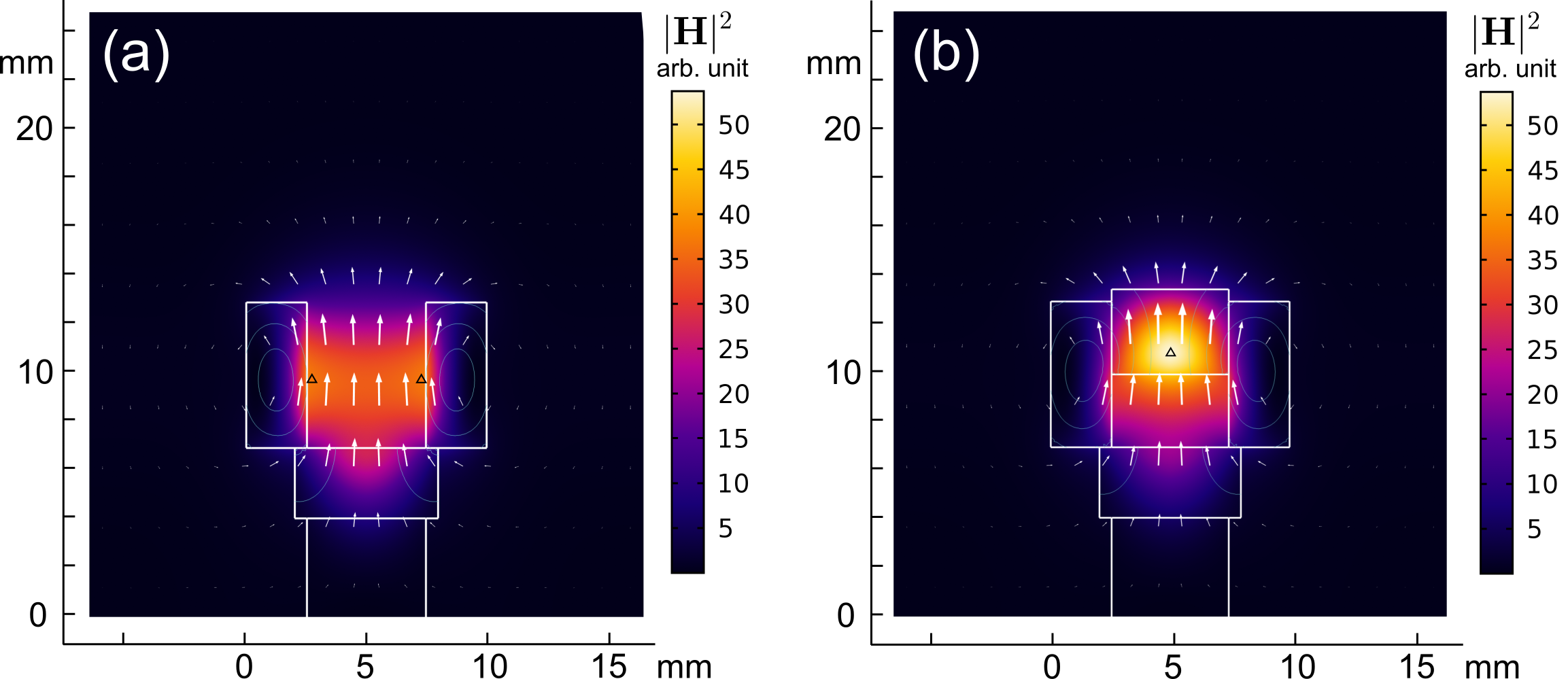}
\caption{\label{fig:vmode}COMSOL simulations of the magnetic field strength $|\textbf{H}|^2$ for the sapphire resonator with (a) no wedge and (b) with wedge. The axes denote the physical dimensions of the inside of the copper cavity, the sapphire ring, stand and wedge. The white borders define the shapes of the ring, stand and wedge}
\end{figure*}

Figure~S~\ref{fig:vmode} shows COMSOL simulations of the magnetic field strength for the sapphire dielectric with no sapphire wedge and with a sapphire wedge (approximated to a cylinder) inserted into the centre. The perpendicular and parallel dielectric constants used for sapphire were 9.3 and 11.5 respectively. Figure~S~\ref{fig:vmode}(b) is identical to Figure~1(d) in the main text.

The region of maximum field strength (maximum $|\textbf{H}|^2$) for Figure~S~\ref{fig:vmode}(a), indicated by black triangles, occurs on the sides of the hole in the sapphire ring. Conversely, Figure~S~\ref{fig:vmode}(b) shows the magnetic field is strongest at the centre within the `wedge', alongside high strength below the wedge where the sample would be placed. For masers and electron paramagnetic resonance (EPR) spectroscopy, it is best to place the sample in the region of highest magnetic field strength for the highest interaction, and so the `wedge' may be advantageously concentrating the field to where the sample is, rather than the field spreading to maxima on the sides of the hole as in Figure~S~\ref{fig:vmode}(a)

The mode volume is defined as $V_{\text{mode}}=\frac{\int_{\text{mode}}|\textbf{H}|^2\ dV}{\text{max}[|\textbf{H}|^2]}$, where \textbf{H} is the magnetic field strength, $\int_{\text{mode}}...\ dV$ indicates integration over and around the entire mode bright spot, and $\text{max}[...]$ denotes the maximum value (scalar) of the functional argument. These integrals can be easily calculated using COMSOL, giving a lower $V_{\text{mode}}$ for Figure~S~\ref{fig:vmode}(b) at 0.18 cm$^{3}$, compared to 1.6 cm$^{3}$ for Figure~S~\ref{fig:vmode}(a). This indicates that the presence of the wedge seems to have shrunk the mode volume as well, which is advantageous for masing.

\section{Mechanical drawings of cavity}
Drawings of the physical dimensions of the cavity are included on the next page. The cavity is made of three parts; part A is the main body which houses the sapphire dielectric, part B is the cap that is screwed onto part A, and part C is the tuning screw which adjusts the height of the inner ceiling for resonator frequency tuning.
\bibliography{Bsupp}
\newpage

\begin{figure*}[pbth]
\includegraphics[width=.75\paperwidth]{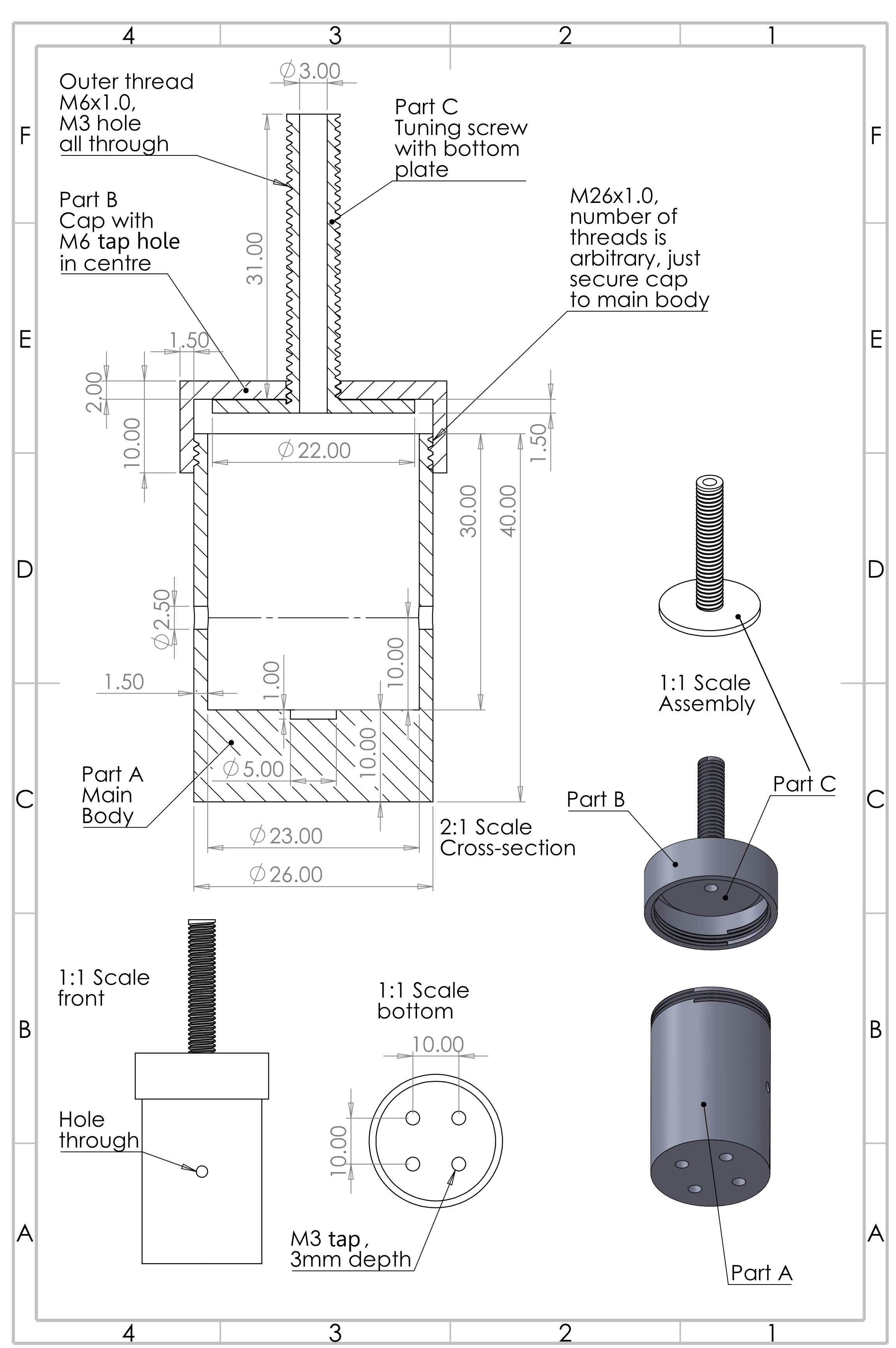}

\end{figure*}